\documentclass{sigchi}

\pdfoutput=1  

\usepackage{newtxtext}
\usepackage{newtxmath}
\usepackage{CJKutf8}

\pdfmapfile{=newtx.map}

\usepackage{amsfonts}
\usepackage{amsmath}
\usepackage{balance}
\usepackage{bbm}
\usepackage{bm}
\usepackage{booktabs}
\usepackage{color}
\usepackage{enumerate}
\usepackage{graphicx}
\usepackage{grffile}      
\usepackage[T1]{fontenc}  
\usepackage[final,stretch=10,spacing=true]{microtype}
\usepackage{multirow}
\usepackage{outlines}
\usepackage{pbox}
\usepackage{spverbatim}
\usepackage[tight]{subfigure}
\usepackage{tabulary}
\usepackage{textgreek}

\usepackage{hyperref}

\newif\ifanonymize    \anonymizefalse
\newif\ifcameraready  \camerareadyfalse
\ifcameraready
  \anonymizefalse
\fi
\ifanonymize
  \usepackage{censor}
\else
  \newcommand{\censor}[1]{#1}
  \newcommand{\blackout}[1]{#1}
\fi

\ifcameraready

\toappear{Permission to make digital or hard copies of part or all of this
  work for personal or classroom use is granted without fee provided that
  copies are not made or distributed for profit or commercial advantage and
  that copies bear this notice and the full citation on the first page.
  Copyrights for third-party components of this work must be honored. For all
  other uses, contact the owner/author(s). Copyright is held by the
  author/owner(s).
  {\emph{CSCW 2014}}, February 15--19, 2014, Baltimore, Maryland, USA. \\
  ACM 978-1-4503-2540-0/14/02. LA-UR 13-23557.\\
  \url{http://dx.doi.org/10.1145/2531602.2531607}}
\else
  \toappear{Preprint version 3. LA-UR 13-23557. Please cite the published
    version in the ACM Digital Library
    (\url{http://dx.doi.org/10.1145/2531602.2531607}).}
\fi

\newcommand{\algname}[1]{\textit{#1}}

\newcommand{\argmin}{\mathop{\rm argmin}}

\newcommand{\doi}[1]{\href{http://dx.doi.org/#1}{\nolinkurl{#1}}}
\newcommand{\etal}{et~al.}

\newcommand{\inhead}[1]{\textbf{#1}}

\newcommand{\mc}{\multicolumn}
\newcommand{\mr}{\multirow}
\newcommand{\px}{\ensuremath{\circ}}
\newcommand{\pv}{\ensuremath{\ast}}
\newcommand{\poi}{\ensuremath{\ast\ast}}
\newcommand{\pooi}{\ensuremath{\text{\ensuremath{\ast}\ensuremath{\ast}\ensuremath{\ast}}}}
\newcommand{\vocab}[1]{\emph{#1}}  

\newcommand{\dy}{\,\mathrm{d}y}
\newcommand{\oc}[1]{\ensuremath{\text{OC}_{#1}}}
\newcommand{\pra}[1]{\ensuremath{\text{PRA}_{#1}}}
\newcommand{\shrunkenfont}{}
\newcommand{\tablefont}{}
\newcommand{\token}[1]{\textit{#1}}

\newcommand{\xcaptionplace}{}
\newcommand{\xcaption}[1]{
  \renewcommand{\xcaptionplace}{\caption{\shrunkenfont{}#1}}
}

\newlength{\spacehackfig}
\newlength{\spacehacktab}
\makeatletter
\def\url@mystyle{%
\@ifundefined{selectfont}{\def\UrlFont{\rm}}{\def\UrlFont{\rmfamily}}}
\g@addto@macro{\UrlNoBreaks}{\do:}
\def\url@mybigstyle{%
\@ifundefined{selectfont}{\def\UrlFont{\rm}}{\def\UrlFont{\large\rmfamily}}}
\g@addto@macro{\UrlNoBreaks}{\do:}
\makeatother
\hypersetup{
  bookmarksnumbered,
  colorlinks=true,
  citecolor=black,
  filecolor=black,
  linkcolor=black,
  urlcolor=black,
  breaklinks=true,
}

\begin{document}

\newcommand{\thetitle}{Inferring the Origin Locations of Tweets\\ with
  Quantitative Confidence}
\title{\thetitle}
\newcommand{\authorreid}{Reid Priedhorsky}
\newcommand{\authorsara}{Sara Y. Del Valle}
\newcommand{\authoraron}{Aron Culotta}

\numberofauthors{2}
\newlength{\myauwidth}
\setlength{\myauwidth}{170pt}
\author{
  \urlstyle{mybig}
  \begin{tabular}[t]{cc}
    \multicolumn{2}{c}{\censor{\authorreid,$^{\boldsymbol{*}}$}
                       \censor{\authoraron,$^{\boldsymbol{\dagger}}$}
                       \censor{\authorsara$^{\boldsymbol{*}}$}}
    \\
    \vspace{-4pt}
    \\
    \parbox{\myauwidth}{
      \centering
      \censor{\affaddr{$^*$Los Alamos National Laboratory}}\\
      \vspace{1pt}
      \censor{\affaddr{Los Alamos, NM}}\\
      \vspace{1pt}
      \censor{\url{{reidpr,sdelvall}@lanl.gov}}}
    &
    \parbox{\myauwidth}{
      \centering
      \censor{\affaddr{$^\dagger$Illinois Institute of Technology}}\\
      \vspace{1pt}
      \censor{\affaddr{Chicago, IL}}\\
      \vspace{1pt}
      \censor{\url{aculotta@iit.edu}}}
  \end{tabular}
}

\ifanonymize
\else
  \hypersetup{
    pdftitle={\thetitle},
    pdfauthor={\authorreid, \authoraron, \authorsara}
  }
\fi

\maketitle

\begin{abstract}


  Social Internet content plays an increasingly critical role in many domains,
  including public health, disaster management, and politics. However, its
  utility is limited by missing geographic information; for example, fewer
  than 1.6\% of Twitter messages (\vocab{tweets}) contain a geotag. We propose
  a scalable, content-based approach to estimate the location of tweets using
  a novel yet simple variant of gaussian mixture models. Further, because
  real-world applications depend on quantified uncertainty for such estimates,
  we propose novel metrics of accuracy, precision, and calibration, and we
  evaluate our approach accordingly. Experiments on 13~million global,
  comprehensively multi-lingual tweets show that our approach yields reliable,
  well-calibrated results competitive with previous computationally intensive
  methods. We also show that a relatively small number of training data are
  required for good estimates (roughly 30,000 tweets) and models are quite
  time-invariant (effective on tweets many weeks newer than the training set).
  Finally, we show that toponyms and languages with small geographic footprint
  provide the most useful location signals.

\end{abstract}

\section{Introduction}

Applications in public health~\cite{dredze_how_2012},
politics~\cite{savage_twitter_2011}, disaster
management~\cite{mcclendon_leveraging_2012}, and other domains are
increasingly turning to social Internet data to inform policy and intervention
strategies. However, the value of these data is limited because the geographic
origin of content is frequently unknown. Thus, there is growing interest in
the task of \vocab{location inference}: given an item, estimate its geographic
\vocab{true origin}.


\xcaption{A tweet originating near Los Angeles, CA. We show the true origin (a
  blue star) and a heat map illustrating the density function that makes up
  our method's estimate. This estimate, whose accuracy was at the 80th
  percentile, was driven by two main factors. The unigram \textit{ca} from the
  location field, visible as the large density oval along the California
  coast, contributed about 12\% of the estimate, while \textit{angeles ca},
  the much denser region around Los Angeles, contributed 87\%. The
  contribution of four other n-grams (\textit{angeles}, \textit{los angeles},
  \textit{obama}, and \textit{los}) was negligible.}
\begin{figure}
  \includegraphics[width=\columnwidth]{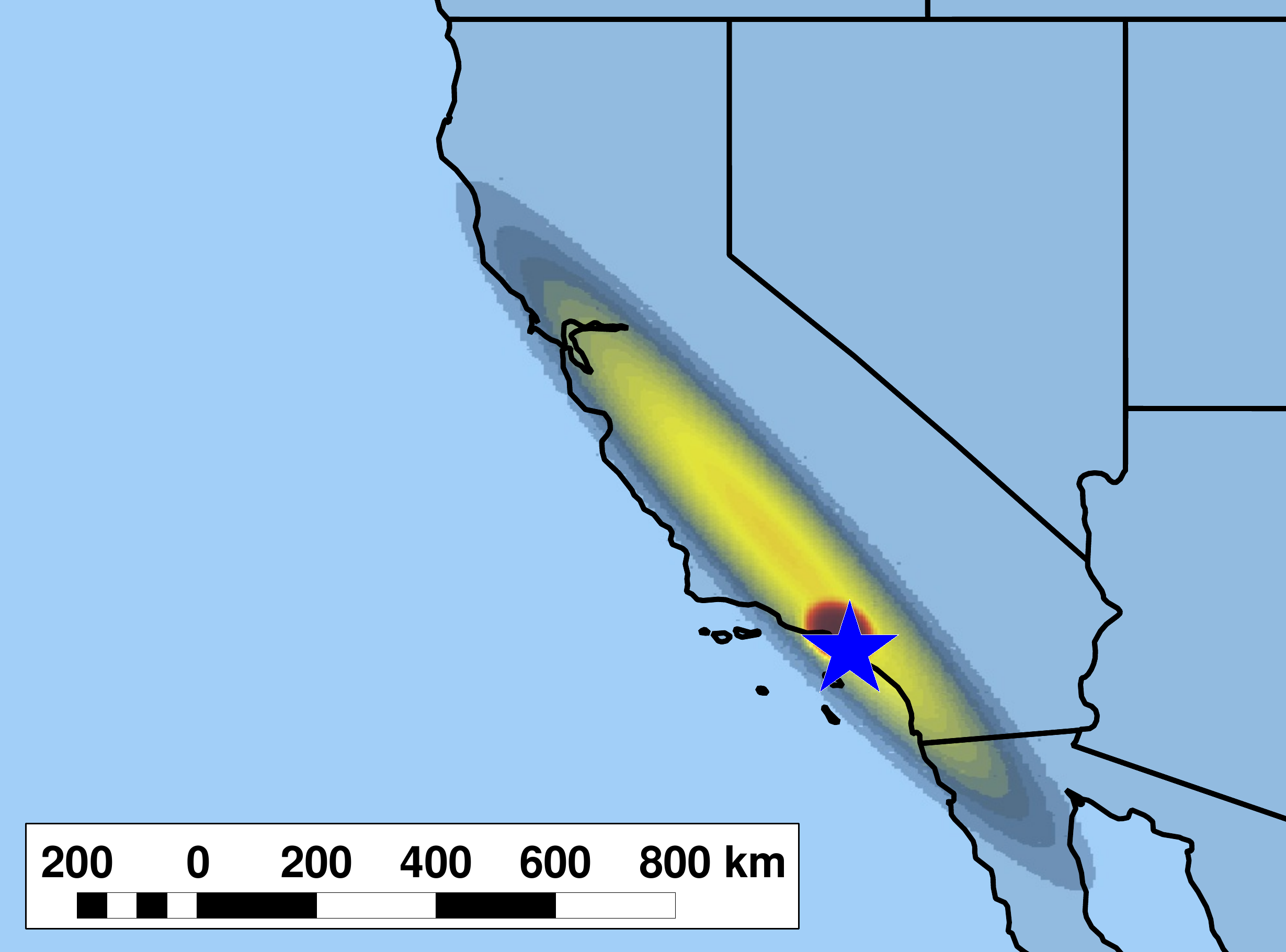}
  \newcommand{\tv}[1]{#1}
  \small
  \begin{tabulary}{\columnwidth}{@{}l@{\hskip 8pt}L@{}}
    \vspace{-7pt} \\
    \textbf{text:} & \tv{Americans are optimistic about the economy \& like what Obama is doing. What is he doing? Campaigning and playing golf? Ignorance is bliss} \\
    \textbf{language:} & \tv{en} \\
    \textbf{location:} & \tv{Los Angeles, CA} \\
    \textbf{time zone:} & \tv{pacifictimeuscanada} \\
  \end{tabulary}
  \vspace{2pt}
  \xcaptionplace
  \label{fig.tweetwin}
\end{figure}

We propose an inference method based on gaussian mixture models
(GMMs)~\cite{mclachlan_finite_2005}. Our models are trained on geotagged
tweets, i.e., messages with user profile and geographic true origin
points.\footnote{Our implementation is open source:
  \censor{\url{http://github.com/reidpr/quac}}} For each unique n-gram, we fit
a two-dimensional GMM to model its geographic distribution. To infer the
origin of a new tweet, we combine previously trained GMMs for the n-grams it
contains, using weights inferred from data; Figure~\ref{fig.tweetwin} shows an
example estimate. This approach is simple, scalable, and competitive with more
complex approaches.

Location estimates using any method contain uncertainty, and it is important
for downstream applications to quantify this uncertainty. While previous work
considers only point estimates, we argue that a more useful form consists of a
density estimate (of a probability distribution) covering the entire globe,
and that estimates should be assessed on three independent dimensions of
accuracy, precision, and calibration. We propose new metrics for doing so.

To validate our approach, we performed experiments on twelve months of tweets
from across the globe, in the context of answering four research questions:

\begin{enumerate}[RQ1.]

\item \inhead{Improved approach.} How can the origin locations of social
  Internet messages be estimated accurately, precisely, and with quantitative
  uncertainty? Our novel, simple, and scalable GMM-based approach produces
  well-calibrated estimates with a global mean accuracy error of roughly
  1,800\,km and precision of 900,000 square kilometers (or better); this is
  competitive with more complex approaches on the metrics available in prior
  work.

\item \inhead{Training size.} How many training data are required? We find
  that approximately 30,000 tweets (i.e., roughly 0.01\% of total daily
  Twitter activity) are sufficient for high-quality models, and that
  performance can be further improved with more training data at a cost of
  increased time and memory. We also find that models are improved by
  including rare n-grams, even those occurring just 3 times.

\item \inhead{Time dependence.} What is the effect of a temporal gap between
  training and testing data? We find that our models are nearly independent of
  time, performing just 6\% worse with a gap of 4~months (vs.\ no gap).

\item \inhead{Location signal sources.} Which types of content provide the
  most valuable location signals? Our results suggest that the user location
  string and time zone fields provide the strongest signals, tweet text and
  user language are weaker but important to offer an estimate for all test
  tweets, and user description has essentially no location value. Our results
  also suggest that mentioning toponyms (i.e., names of places), especially at
  the city scale, provides a strong signal, as does using languages with a
  small geographic footprint.

\end{enumerate}

The remainder of our paper is organized as follows. We first survey related
work, then propose desirable properties of a location inference method and
metrics which measure those properties. We then describe our experimental
framework and detail our mixture model approach. Finally, we discuss our
experimental results and their implications. Appendices with implementation
details follow the body of the paper.


\section{Related Work}
\label{sec.related}

Over the past few years, the problem of inferring the origin locations of
social Internet content has become an increasingly active research area.
Below, we summarize the four primary lines of work and contrast them with this
paper.

\subsection{Geocoding}

Perhaps the simplest approach to location inference is \vocab{geocoding}:
looking up the user profile's free-text location field in a \vocab{gazetteer}
(list of toponyms), and if a match is found, inferring that the message
originated from the matching place. Researchers have used commercial geocoding
services such as Yahoo!\ Geocoder~\cite{valkanas_location_2012},
U.S.\ Geological Survey data~\cite{paradesi_geotagging_2011}, and
Wikipedia~\cite{hecht_tweets_2011} to do this. This technique can be extended
to the message text itself by first using a \vocab{geoparser} named-entity
recognizer to extract toponyms~\cite{gelernter_geo-parsing_2011}.

Schulz \etal~\cite{schulz_multi-indicator_2013} recently reported accurate
results using a scheme which combines multiple geocoding sources, including
Internet queries. Crucial to its performance was the discovery that an
additional 26\% of tweets can be matched to precise coordinates using text
parsing and by following links to location-based services (FourSquare, Flickr,
etc.), an approach that can be incorporated into competing methods as well.
Another 8\% of tweets -- likely the most difficult ones, as they contain the
most subtle location evidence -- could not be estimated and are not included
in accuracy results.

In addition to one or more accurate, comprehensive gazetteers, these
approaches require careful text cleaning before geocoding is attempted, as
grossly erroneous false matches are common~\cite{hecht_tweets_2011}, and they
tend to favor precision over recall (because only toponyms are used as
evidence). Finally, under one view, our approach essentially infers a
probabilistic gazetteer that weights toponyms (and pseudo-toponyms) according
to the location information they actually carry.

\subsection{Statistical classifiers}

These approaches build a statistical mapping of text to discrete pre-defined
regions such as cities and countries (i.e., treating ``origin location'' as
membership in one of these classes rather than a geographic point); thus, any
token can be used to inform location inference.

We categorize this work by the type of classifier and by place granularity.
For example, Cheng~\etal{} apply a variant of nai\"{v}e Bayes to classify
messages by city~\cite{cheng_you_2010}, Hecht~\etal{} use a similar classifier
at the state and country level~\cite{hecht_tweets_2011}, and Kinsella~\etal{}
use language models to classify messages by neighborhood, city, state, zip
code, and country~\cite{kinsella_``im_2011}. Mahmud~\etal{} classify users by
city with higher accuracy than Cheng~\etal{} by combining a hierarchical
classifier with many heuristics and gazetteers~\cite{mahmud_where_2012}. Other
work instead classifies messages into arbitrary regions of
fixed~\cite{ohare_modeling_2012,wing_simple_2011} or dynamic
size~\cite{roller_supervised_2012}. All of these require aggressively
smoothing estimates for regions with few observations~\cite{cheng_you_2010}

Recently, Chang \etal~\cite{chang_phillies_2012} classified tweet text by
city using GMMs. While more related to the present paper because of the
underlying statistical technique, this work is still fundamentally a
classification approach, and it does not attempt the probabilistic evaluation
that we advocate. Additionally, the algorithm resorts to heuristic feature
selection to handle noisy n-grams; instead, we offer two learning algorithms
to set n-gram weights which are both theoretically grounded and empirically
crucial for accuracy.

Fundamentally, these approaches can only classify messages into regions
specified before training; in contrast, our GMM approach can be used both for
direct location inference as well as classification, even if regions are
post-specified.

\subsection{Geographic topic models}

\begin{sloppypar}
These techniques endow traditional topic models~\cite{blei_latent_2003} with
location awareness~\cite{wang_mining_2007}. Eisenstein~\etal{} developed a
\vocab{cascading topic model} that produces region-specific topics and used
these topics to infer the locations of Twitter
users~\cite{eisenstein_latent_2010}; follow-on work uses \vocab{sparse
  additive models} to combine region-specific, user-specific, and
non-informative topics more
efficiently~\cite{eisenstein_sparse_2011,hong_discovering_2012}.
\end{sloppypar}

Topic modeling does not require explicit pre-specified regions. However,
regions are inferred as a preprocessing step: Eisenstein~\etal{} with a
Dirichlet Process mixture~\cite{eisenstein_latent_2010} and Hong~\etal{} with
K-means clustering~\cite{hong_discovering_2012}. The latter also suggests that
more regions increases inference accuracy.

While these approaches result in accurate models, the bulk of modeling and
computational complexity arises from the need to produce geographically
coherent topics. Also, while topic models can be parallelized with
considerable effort, doing so often requires approximations, and their global
state limits the potential speedup. In contrast, our approach focusing solely
on geolocation is simpler and more scalable.

Finally, the efforts cited restrict messages to either the United States or
the English language, and they report simply the mean and median distance
between the true and predicted location, omitting any precision or uncertainty
assessment. While these limitations are not fundamental to topic modeling, the
novel evaluation and analysis we provide offer new insights into the strengths
and weaknesses of this family of algorithms.

\subsection{Social network information}

Recent work suggests that using social link information (e.g., followers or
friends) can aid in location
inference~\cite{chandra_estimating_2011,davis_jr._inferring_2011}. We view
these approaches as complementary to our own; accordingly, we do not explore
them more deeply at present.

\subsection{Contrasting our approach}

We offer the following principal distinctions compared to prior work:
(a)~location estimates are multi-modal probability distributions, rather than
points or regions, and are rigorously evaluated as such, (b)~because we deal
with geographic coordinates directly, there is no need to pre-specify regions
of interest; (c)~no gazetteers or other supplementary data are required, and
(d)~we evaluate on a dataset that is more comprehensive temporally (one year
of data), geographically (global), and linguistically (all languages except
Chinese, Thai, Lao, Cambodian, and Burmese).

\section{Experiment design}

In this section, we present three properties of a good location estimate,
metrics and experiments to measure them, and new algorithms motivated by them.

\subsection{What makes a good location estimate?}
\label{sec.what}

An estimate of the origin location of a message should be able to answer two
closely related but different questions:

\begin{enumerate}[Q1.]

\item What is the \vocab{true origin} of the message? That is, at which
  geographic point was the person who created the message located when he or
  she did so?

\item Was the true origin within a specified geographical region? For example,
  did a given message originate from Washington State?

\end{enumerate}

It is inescapable that all estimates are uncertain. We argue that they should
be quantitatively treated as such and offer probabilistic answers to these
questions. That is, we argue that a location estimate should be a
\vocab{geographic density estimate}: a function which estimates the
probability of every point on the globe being the true origin. Considered
through this lens, a high-quality estimate has the following properties:

\begin{itemize}

\item It is \vocab{accurate}: the density of the estimate is skewed strongly
  towards the true origin (i.e., the estimate rates points near the true
  origin as more probable than points far from it). Then, Q1 can be answered
  effectively because the most dense regions of the distribution are near the
  true origin, and Q2 can be answered effectively because if the true origin
  is within the specified region, then much of the distribution's density will
  be as well.

\item It is \vocab{precise}: the most dense regions of the estimate are
  compact. Then, Q1 can be answered effectively because fewer candidate
  locations are offered, and Q2 can be answered effectively because the
  distribution's density is focused within few distinct regions.

\item It is \vocab{well calibrated}: the probabilities it claims are close to
  the true probabilities. Then, both questions can be answered effectively
  regardless of the estimate's accuracy and precision, because its uncertainty
  is quantified. For example, the two estimates ``the true origin is within
  New York City with 90\% confidence'' and ``the true origin is within North
  America with 90\% confidence'' are both useful even though the latter is
  much less accurate and precise.

\end{itemize}

Our goal, then, is to discover an \vocab{estimator} which produces estimates
that optimize the above properties.

\subsection{Metrics}
\label{sec.metrics}

We now map these properties to operationalizable metrics. This section
presents our metrics and their intuitive reasoning; rigorous mathematical
implementations are in the appendices.

\subsubsection{Accuracy}

Our core metric to evaluate the accuracy of an estimate is
\vocab{comprehensive accuracy error} (CAE): the expected distance between the
true origin and a point randomly selected from the estimate's density
function, or in other words, the mean distance between the true origin and
every point on the globe, weighted by the estimate's density value.\footnote{A
  similar metric, called Expected Distance Error, has been proposed by
  Cho~\etal{} for a different task of user
  tracking~\cite{cho_friendship_2011}.} The goal here is to offer a notion of
the distance from the true origin to the density estimate as a whole.


\xcaption{True origins of tweets having the unigram \token{washington} in the
  location field of the user's profile.}
\begin{figure}
  \includegraphics[width=\columnwidth]{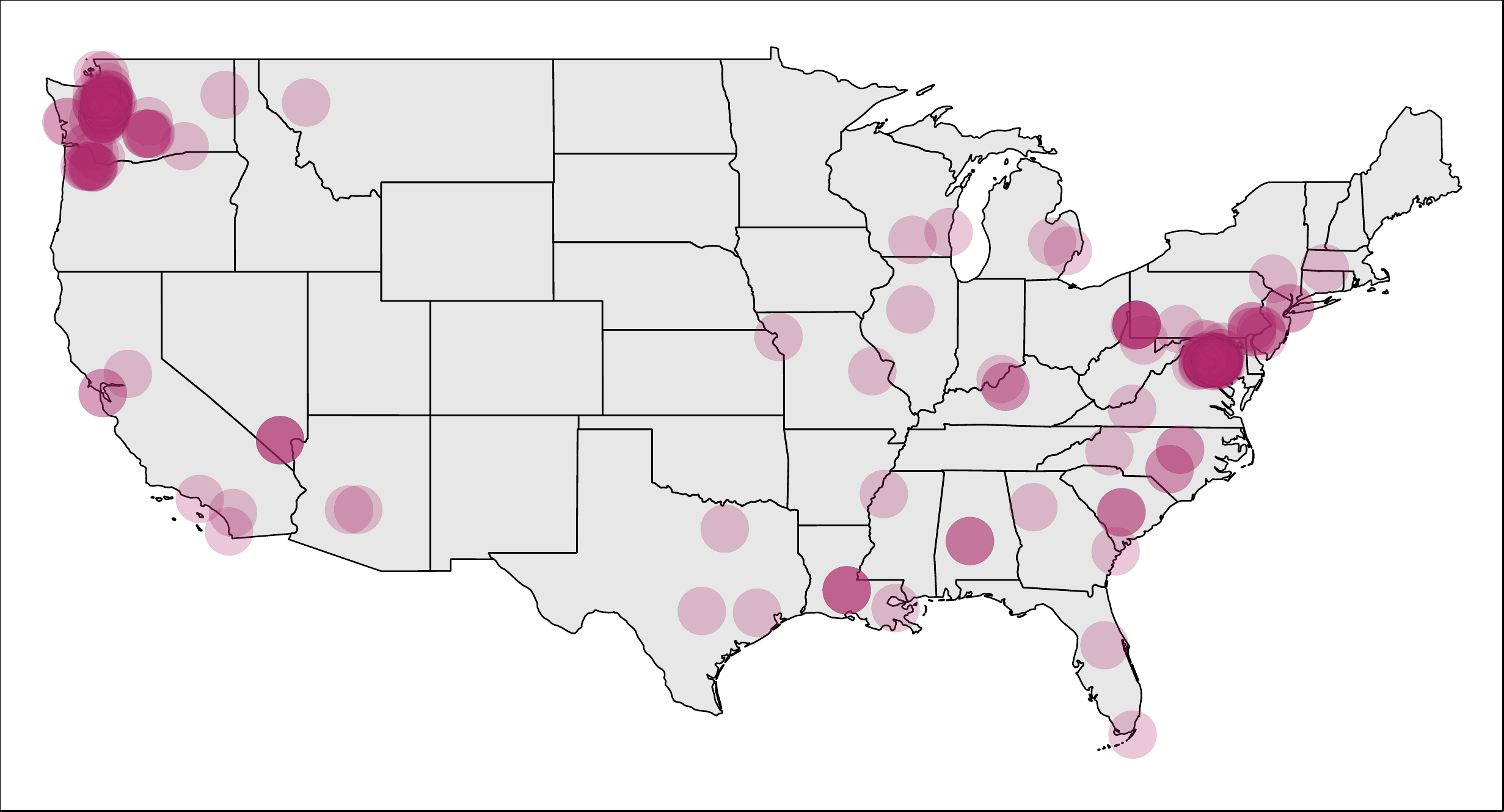}
  \vspace{\spacehackfig}
  \xcaptionplace
  \label{fig.washington}
\end{figure}

This contrasts with a common prior metric that we refer to as \vocab{simple
  accuracy error} (SAE): the distance from the best single-point estimate to
the true origin. Figure~\ref{fig.washington} illustrates this contrast. The
tight clusters around both Washington, D.C. and Washington State suggest that
any estimate based on the unigram \token{washington} is inherently bimodal;
that is, no single point at either cluster or anywhere in between is a good
estimated location. More generally, SAE is a poor match for the continuous,
multi-modal density estimates that we argue are more useful for downstream
analysis, because good single-point distillations are often unavailable.
However, we report both metrics in order to make comparisons with prior work.

The units of CAE (and SAE) are kilometers. For a given estimator (i.e., a
specific algorithm which produces location estimates), we report \vocab{mean
  comprehensive accuracy error} (MCAE), which is simply the mean of each
estimate's CAE. $\text{CAE} \geq 0$, and an ideal estimator has $\text{MCAE} =
0$.

\subsubsection{Precision}

In order to evaluate precision, we extend the notion of one-dimensional
prediction intervals~\cite{brown_confidence_2002,geisser_predictive_1993} to
two dimensions. An estimate's \vocab{prediction region} is the minimal,
perhaps non-contiguous geographic region which contains the true origin with
some specified probability (the region's \vocab{coverage}).

Accordingly, the metric we propose for precision is simply the area of this
region: \vocab{prediction region area} (PRA) parameterized by the coverage,
e.g., \pra{50} is the area of the minimal region which contains the true
origin with 50\% probability.

Units are square kilometers. For a given estimator, we report \vocab{mean
  prediction region area} (MPRA), i.e., the mean of each estimate's PRA.
$\text{PRA} \ge 0$; an ideal estimator has $\text{MPRA} = 0$.

\subsubsection{Calibration}

Calibration is tested by measuring the difference between an estimate's
claimed probability that a particular point is the true origin and its actual
probability.

We accomplish this by building upon prediction regions. That is, given a set
of estimates, we compute a prediction region at a given coverage for each
estimate and measure the fraction of true origins that fall within the
regions. The result should be close to the specified coverage. For example,
for prediction regions at coverage 0.5, the fraction of true origins that
actually fall within the prediction region should be close to 0.5.

We refer to this fraction as \vocab{observed coverage} (OC) at a given
expected coverage; for example, \oc{50} is the observed coverage for an
expected coverage of 0.5. (This measure is common in the statistical
literature for one-dimensional problems~\cite{brown_confidence_2002}.)
Calibration can vary among different expected coverage levels (because fitted
density distributions may not exactly match actual true origin densities), so
multiple coverage levels should be reported (in this paper, \oc{50} and
\oc{90}).

Note that \oc{} is defined at the estimator level, not for single messages.
\oc{} is unitless, and $0 \le \oc{} \le 1$. An ideal estimator has observed
coverage equal to expected coverage, an overconfident estimator has observed
less than expected, and an underconfident one greater.

\subsection{Experiment implementation}
\label{sec.expt}

In this section, we explain the basic structure of our experiments: data
source, preprocessing and tokenization, and test procedures.

\subsubsection{Data}
\label{sec.data}


We used the Twitter Streaming API to collect an approximately continuous 1\%
sample of all global tweets from January 25, 2012 to January 23, 2013. Between
0.8\% and 1.6\% of these, depending on timeframe, contained a geotag (i.e.,
specific geographic coordinates marking the true origin of the tweet, derived
from GPS or other automated means), yielding a total of approximately 13
million geotagged tweets.\footnote{As in prior
  work~\cite{eisenstein_latent_2010, hong_discovering_2012,
    roller_supervised_2012}, we ignore the sampling bias introduced by
  considering only geotagged tweets. A preliminary analysis suggests this bias
  is limited. In a random sample of 11,694,033 geotagged and 17,175,563
  non-geotagged tweets from 2012, we find a correlation of 0.85 between the
  unigram frequency vectors for each set; when retweets are removed, the
  correlation is 0.93.}

We tokenized the message text (\vocab{tx}), user description (\vocab{ds}), and
user location (\vocab{lo}) fields, which are free-text, into bigrams by
splitting on Unicode character category and script boundaries and then further
subdividing bigrams appearing to be Japanese using the TinySegmenter
algorithm~\cite{hagiwara_tinysegmenter_????}.\footnote{More complex
  tokenization methods yielded no notable effect.} This covers all languages
except a few that have low usage on Twitter: Thai, Lao, Cambodian, and Burmese
(which do not separate words with a delimiter) as well as Chinese (which is
difficult to distinguish from Japanese). For example, the string
``Can't wait for
\begin{CJK}{UTF8}{min}私の\end{CJK}'' becomes the set of
bigrams \emph{can}, \emph{t}, \emph{wait}, \emph{for},
\begin{CJK}{UTF8}{min}私\end{CJK},
\begin{CJK}{UTF8}{min}の\end{CJK}, \emph{can t}, \emph{t wait},
\emph{wait for}, \emph{for} \begin{CJK}{UTF8}{min}私\end{CJK}, and
\begin{CJK}{UTF8}{min}私\end{CJK}
\begin{CJK}{UTF8}{min}の\end{CJK}.
(Details of our algorithm are presented in the appendices.)

For the language (\vocab{ln}) and time zone (\vocab{tz}) fields, which are
selected from a set of options, we form n-grams by simply removing whitespace
and punctuation and converting to lower-case. For example, ``Eastern Time (US
\& Canada)'' becomes simply \emph{easterntimeuscanada}.


\subsubsection{Experiments}

Each experiment is implemented using a Python script on tweets selected with a
regular schedule. For example, we might train a model on all tweets from May~1
and test on a random sample of tweets from May~2, then train on May~7 and test
on May~8, etc. This schedule has four parameters:

\begin{itemize}

\item \inhead{Training duration.} The length of time from which to select
  training tweets. We used all selected tweets for training, except only the
  first tweet from a given user is retained, to avoid over-weighting frequent
  tweeters.

\item \inhead{Test duration.} The length of time from which to select test
  tweets. In all experiments, we tested on a random sample of 2,000 tweets
  selected from one day. We excluded users with a tweet in the training set
  from testing, in order to avoid tainting the test set.

\item \inhead{Gap.} The length of time between the end of training data and
  the beginning of test data.

\item \inhead{Stride.} The length of time from the beginning of one training
  set to the beginning of the next. This was fixed at 6 days unless otherwise
  noted.

\end{itemize}

For example, an experiment with training size of one day, no gap, and stride
of 6 days would schedule 61 tests across our 12 months of data and yield
results which were the mean of the 58 tests with sufficient data (i.e., 3
tests were not attempted due to missing data). The advantage of this approach
is that test data always chronologically follow training data, minimizing
temporal biases and better reflecting real-world use.

We built families of related experiments (as described below) and report
results on these families.

\section{Our approach: Geographic GMMs}
\label{sec.gmm}

Here, we present our location inference approach. We first motivate and
summarize it, then detail the specific algorithms we tested. (Mathematical
implementations are in the appendices.)

\subsection{Motivation}

Examining the geographic distribution of n-grams can suggest appropriate
inference models. For example, recall Figure~\ref{fig.washington} above; the
two clusters, along with scattered locations elsewhere, suggest that a
multi-modal distribution consisting of two-dimensional gaussians may be a
reasonable fit.

Based on this intuition and coupled with the desiderata above, we propose an
estimator using one of the mature density estimation techniques:
\vocab{gaussian mixture models} (GMMs). These models are precisely the
weighted sum of multiple gaussian (normal) distributions and have natural
probabilistic interpretations. Further, they have previously been applied to
human mobility
patterns~\cite{cho_friendship_2011,gonzalez_understanding_2008}.

Our algorithm is summarized as follows:

\begin{enumerate}

\item For each n-gram that appears more than a threshold number of times in the
  training data, fit a GMM to the true origin points of the tweets in the
  training set that contain that n-gram. This n-gram/GMM mapping forms the
  trained location model.

\item To locate a test tweet, collect the GMMs from the location models which
  correspond to n-grams in the test tweet. The weighted sum of these GMMs ---
  itself a GMM --- is the geographic density function which forms the estimate
  of the test tweet's location.

\end{enumerate}

It is clear that some n-grams will carry more location information than
others. For example, n-gram density for the word \token{the} should have high
variance and be dispersed across all English-speaking regions; on the other
hand, density for \token{washington} should be concentrated in places named
after that president.\footnote{Indeed, Eisenstein~\etal{} attribute the poor
  performance of several of their baselines to this tendency of uninformative
  words to dilute the predictive power of informative
  words~\cite{eisenstein_latent_2010}.} That is, n-grams with much location
information should be assigned high weight, and those with little information
low weight --- but not zero, so that messages with only low-information
n-grams will have a quantifiably poor estimate rather than none at all.
Accordingly, we propose three methods to set the GMM weights.

\subsection{Weighting by quality properties}
\label{sec.weight-quality}

One approach is to simply assign higher weight to GMMs which have a crisper
signal or fit the data better. We tested 15 \vocab{quality properties} which
measure this in different ways.

We tried weighting each GMM by the inverse of (1)~the number of fitted points,
(2)~the spatial variance of these points, and (3)~the number of components in
the mixture. We also tried metrics based on the covariance matrices of the
gaussian components: the inverse of (4)~the sum of all elements, and (5)~the
sum of the products of the elements in each matrix. Finally, we tried
normalizing: by both the number of fitted points (properties 6--9) and the
number of components (10--13). Of these, property 5, which we call
\algname{GMM-Qpr-Covar-Sum-Prod}, performed the best, so we carry it forward
for discussion.

Additionally, we tried two metrics designed specifically to test goodness of
fit: (14) Akaike information criterion~\cite{akaike_new_1974} and (15)
Bayesian information criterion~\cite{schwarz_estimating_1978}, transformed
into weights by subtracting from the maximum observed value. Of this pair,
property 14, which we call \algname{GMM-Qpr-AIC}, performed best, so we carry
it forward.

\subsection{Weighting by error}
\label{sec.weight-error}

Another approach is to weight each n-gram by its error among the training set.
Specifically, for each n-gram in the learned model, we compute the error of its
GMM (CAE or SAE) against each of the points to which it was fitted. We then
raise this error to a power (in order to increase the dominance of relatively
good n-grams over relatively poor ones) and use the inverse of this value as
the n-gram's weight (i.e., larger errors yield smaller weights).

We refer to these algorithms as (for example) \algname{GMM-Err-SAE4}, which
uses the SAE error metric and an exponent of 4. We tried exponent values from
0.5 to 10 as well as both CAE and SAE; because the latter was faster and gave
comparable results, we report only SAE.

\subsection{Weighting by optimization}
\label{sec.weight-optimization}

The above approaches are advantaged by varying degrees of speed and
simplicity. However, it seems plausibly better to learn optimized weights from
the data themselves. Our basic approach is to assign each n-gram a set of
features with their own weights, let each n-gram's weight be a linear
combination of the feature weights, and use gradient descent to find feature
weights such that the total error across all n-grams is minimized (i.e., total
geo-location accuracy is maximized).

For optimization, we tried three types of n-gram features:

\begin{enumerate}

\item The quality properties noted above (\algname{Attr}).

\item Identity features. That is, the first n-gram had Feature~1 and no others,
  the second n-gram had Feature~2 and no others, and so on (\algname{ID}).

\item Both types of features (\algname{Both}).

\end{enumerate}

Finally, we further classify these algorithms by whether we fit a mixture for
each n-gram (\algname{GMM}) or a single gaussian (\algname{Gaussian}). For
example, \algname{GMM-Opt-ID} uses GMMs and weights optimized using ID
features only.

\subsection{Baseline weighting algorithms}

As two final baselines, we considered \algname{GMM-All-Tweets}, which fits a
single GMM to all tweets in the training set and returns that GMM for all
locate operations, and \algname{GMM-One}, which weights all n-gram mixtures
equally.

\section{Results}
\label{sec.results}

We present in this section our experimental results and discussion, framed in
the context of our four research questions. (In addition to the experiments
described in detail above, we tried several variants that had limited useful
impact. These results are summarized in the appendices.)

\subsection{RQ1: Improved approach}
\label{sec.rq1}

Here we evaluate the performance of our algorithms, first with a comparison
between each other and then against prior work (which is less detailed due to
available metrics).

%
%
\xcaption{Performance of key algorithms; we report the mean and standard
  deviation of each metric across each experiment's tests. MCAE and MSAE are
  in kilometers, M\pra{\bm{50}} is in thousands of $\text{km}^\text{2}$, and
  $\oc{\bm{\beta}}$ is unitless. RT is the mean run time, in minutes, of one
  train-test cycle using 8 threads on 6100-series Opteron processors running
  at 1.9\,GHz.}
\begin{table*}[t]
  \centering
  \tablefont
  \begin{tabular}{@{}lr@{\,$\pm$\,}rr@{\,$\pm$\,}rr@{\,$\pm$\,}rr@{\,$\pm$\,}rr@{\,$\pm$\,}rr@{}}
    \toprule
      \mc{1}{@{}c}{\textbf{Algorithm}}
    & \mc{2}{c}{\textbf{MCAE}}
    & \mc{2}{c}{\textbf{MSAE}}
    & \mc{2}{c}{$\mathbf{MPRA}_{\bm{50}}$}
    & \mc{2}{c}{$\mathbf{OC}_{\bm{50}}$}
    & \mc{2}{c}{$\mathbf{OC}_{\bm{90}}$}
    & \mc{1}{c@{}}{\textbf{RT}}\\
    \cmidrule(r){1-1}
    \cmidrule(lr){2-5}
    \cmidrule(lr){6-7}
    \cmidrule(lr){8-11}
    \cmidrule(l){12-12}
    GMM-Err-SAE10 & 1735 & 81 & 1510 & 76 & 824 & 75.8 & 0.453 & 0.012 & 0.724 & 0.013 & 10.6 \\
GMM-Err-SAE4 & 1826 & 82 & 1565 & 78 & 934 & 69.9 & 0.497 & 0.012 & 0.775 & 0.013 & 10.6 \\
GMM-Opt-ID & 1934 & 77 & 1578 & 67 & 1661 & 171.0 & 0.584 & 0.017 & 0.864 & 0.011 & 29.8 \\
GMM-Err-SAE2 & 2173 & 82 & 1801 & 76 & 1192 & 92.5 & 0.567 & 0.012 & 0.848 & 0.011 & 11.0 \\
GMM-Qpr-Covar-Sum-Prod & 2338 & 123 & 2084 & 115 & 1337 & 123.1 & 0.485 & 0.013 & 0.736 & 0.013 & 9.1 \\
Gaussian-Opt-ID & 2445 & 81 & 1635 & 69 & 6751 & 377.5 & 0.731 & 0.015 & 0.902 & 0.011 & 30.2 \\
GMM-Opt-Both & 4780 & 506 & 4122 & 469 & 4207 & 811.2 & 0.796 & 0.078 & 0.943 & 0.052 & 23.4 \\
GMM-Opt-Attr & 4803 & 564 & 4146 & 505 & 4142 & 811.4 & 0.801 & 0.079 & 0.947 & 0.053 & 22.4 \\
GMM-One & 5147 & 221 & 4439 & 251 & 4235 & 443.9 & 0.852 & 0.013 & 0.982 & 0.003 & 10.0 \\
GMM-Qpr-AIC & 5154 & 226 & 4454 & 252 & 4249 & 474.9 & 0.851 & 0.013 & 0.982 & 0.003 & 10.0 \\
GMM-All-Tweets & 7871 & 156 & 7072 & 210 & 5243 & 882.7 & 0.480 & 0.020 & 0.900 & 0.012 & 15.5 \\

    \bottomrule
  \end{tabular}
  \vspace{\spacehacktab}
  \xcaptionplace
  \label{tab.weights}
\end{table*}

\subsubsection{Performance of our algorithms}

We tested each of our algorithms with one day of training data and no gap, all
fields except user description, and minimum n-gram instances set to 3
(detailed reasoning for these choices is given below in further experiments).
With a stride of 6 days, this yielded 58 tests on each algorithm, with 3 tests
not attempted due to gaps in the data.  Table~\ref{tab.weights} summarizes our
results, making clear the importance of choosing n-gram weights well.

Considering accuracy (MCAE), \algname{GMM-Err-SAE10} is 10\% better than the
best optimization-based algorithm (\algname{GMM-Opt-ID}) and 26\% better than
the best property-based algorithm (\algname{GMM-Qpr-Covar-Sum-Prod}); the
baselines \algname{GMM-One} and \algname{GMM-All-Tweets} performed poorly.
These results suggest that a weighting scheme directly related to performance,
rather than the simpler quality properties, is important --- even including
quality properties in optimization (\algname{-Opt-Attr} and
\algname{-Opt-Both}) yields poor results. Another highlight is the poor
performance of \algname{Gaussian-Opt-ID} vs.\ \algname{GMM-Opt-ID}. Recall
that the former uses a single Gaussian for each n-gram; as such, it cannot fit
the multi-modal nature of these data well.

Turning to precision (M\pra{50}), the advantage of \algname{GMM-Err-SAE10} is
further highlighted; it is 50\% better than \algname{GMM-Opt-ID} and 38\%
better than \algname{GMM-Qpr-Covar-Sum-Prod} (note that the relative order of
these two algorithms has reversed).

However, calibration complicates the picture. While \algname{GMM-Err-SAE10} is
somewhat overconfident at coverage level 0.5 ($\oc{50} = 0.453$ instead of the
desired 0.5), \algname{GMM-Err-SAE4} is calibrated very well at this level
($\oc{50} = 0.497$) and has better calibration at coverage 0.9 ($\oc{90} =
0.775$ instead of 0.724). \algname{GMM-Opt-ID} has still better calibration at
this level ($\oc{90} = 0.864$), though worse at coverage 0.5 ($\oc{50} =
0.584$), and interestingly it is overconfident at one level and underconfident
at the other. A final observation is that some algorithms with poor accuracy
are quite well calibrated at the 0.9 coverage level
(\algname{Gaussian-Opt-ID}) or both levels (\algname{GMM-All-Tweets}). In
short, our calibration results imply that algorithms should be evaluated at
multiple coverage levels, and in particular gaussians may not be quite the
right distribution to fit.

These performance results, which are notably inconsistent between the three
metrics, highlight the value of carefully considering and tuning all three of
accuracy, precision, and calibration. For the remainder of this paper, we will
focus on \algname{GMM-Err-SAE4}, with its simplicity, superior calibration,
time efficiency, and second-best accuracy and precision.

\subsubsection{Is CAE necessary?}

A plausible hypothesis is that the more complex CAE metric is not needed, and
algorithm accuracy can be sufficiently well judged with the simpler and faster
SAE. However, \algname{Gaussian-Opt-ID} offers evidence that this is not the
case: while it is only 4\% worse than \algname{GMM-Err-SAE4} on MSAE, the
relative difference is nearly 6 times greater in MCAE.

Several other algorithms are more consistent between the two metrics, so SAE
may be appropriate in some cases, but caution should be used, particularly
when comparing different types of algorithms.

\subsubsection{Distribution of error}

\xcaption{Accuracy of each estimate using selected algorithms, in descending
  order of CAE.}
\begin{figure}
  \includegraphics[width=\columnwidth]{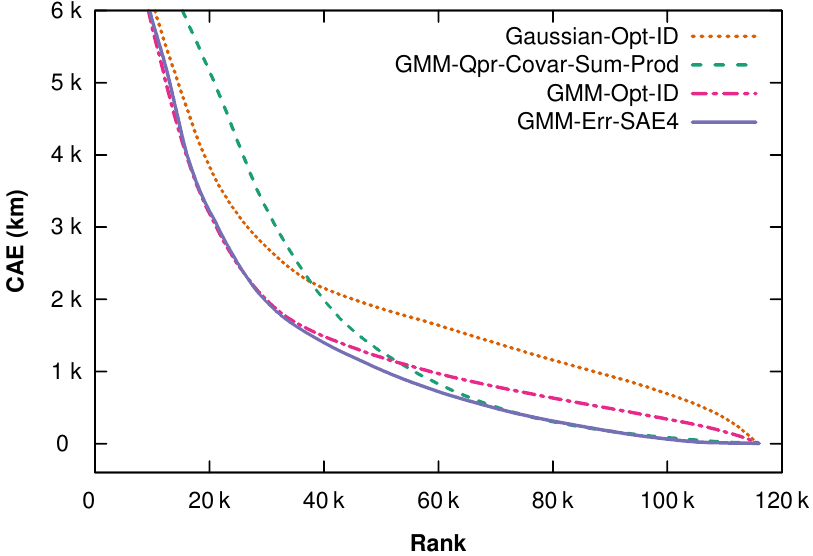}
  \vspace{\spacehackfig}
  \xcaptionplace
  \label{fig.cae}
\end{figure}


Figure \ref{fig.cae} plots the CAE of each estimate from four key algorithms.
These curves are classic long-tail distributions (as are similar ones for
\pra{50} omitted for brevity); that is, a relatively small number of difficult
tweets comprise the bulk of the error. Accordingly, summarizing our results by
median instead of mean may be of some value: for example, the median CAE of
\algname{GMM-Err-SAE4} is 778\,km, and its median \pra{50} is
83,000\,$\text{km}^2$ (roughly the size of Kansas or Austria). However, we
have elected to focus on reporting means in order to not conceal poor
performance on difficult tweets.

It is plausible that different algorithms may perform poorly on different
types of test tweets, though we have not explored this; the implication is
that selecting different strategies based on properties of the tweet being
located may be of value.

\subsubsection{Compared to prior work with the Eisenstein data set}

\xcaption{Our algorithms compared with previous work, using the dataset from
  Eisenstein~\etal~\protect\cite{eisenstein_latent_2010}. The \textit{n-grams}
  column reports the mean number of n-grams used to locate each test tweet.}
\begin{table}
  \centering
  \tablefont
  \begin{tabular}{@{}lrrrr@{}}
    \toprule
      \mc{1}{@{}c}{\mr{2}{*}{\textbf{Algorithm}}}
    & \mc{2}{c}{\textbf{SAE}}
    \\
    & \mc{1}{c}{\textbf{Mean}}
    & \mc{1}{c}{\textbf{Median}}
    & \mc{1}{c}{\textbf{\oc{\bm{50}}}}
    & \mc{1}{c@{}}{\textbf{n-grams}}\\
    \cmidrule(r){1-1}
    \cmidrule(lr){2-3}
    \cmidrule(lr){4-4}
    \cmidrule(l){5-5}
    Hong~\etal~\cite{hong_discovering_2012}        &     & 373 & & \\
    Eisenstein~\etal~\cite{eisenstein_sparse_2011} & 845 & 501 & & \\
    \textbf{GMM-Opt-ID}                            & \textbf{870} & \textbf{534} & \textbf{0.50} & \textbf{19}\\
    Roller~\etal~\cite{roller_supervised_2012}     & 897 & 432 & & \\
    Eisenstein~\etal~\cite{eisenstein_latent_2010} & 900 & 494 & & \\
    \textbf{GMM-Err-SAE6}                          & \textbf{946} & \textbf{588} & \textbf{0.50} & \textbf{153}\\
    \textbf{GMM-Err-SAE16}                         & \textbf{954} & \textbf{493} & \textbf{0.36} & \textbf{37}\\
    Wing~\etal~\cite{wing_simple_2011}             & 967 & 479 & & \\
    \textbf{GMM-Err-SAE4}                          & \textbf{985} & \textbf{684} & \textbf{0.55} & \textbf{182}\\
    \bottomrule
  \end{tabular}
  \vspace{\spacehacktab}
  \xcaptionplace
  \label{tab.cmu}
\end{table}

Table~\ref{tab.cmu} compares \algname{GMM-Opt-ID} and \algname{GMM-Err-SAE} to
five competing approaches using data from
Eisenstein~\etal~\cite{eisenstein_latent_2010}, using mean and median SAE (as
these were the only metrics reported).

These data and our own have important differences. First, they are limited to
tweets from the United States --- thus, we expect lower error here than in our
data, which contain tweets from across the globe. Second, these data were
created for user location inference, not message location (that is, they are
designed for methods which assume users tend to stay near the same location,
whereas our model makes no such assumption and thus may be more appropriate
when locating messages from unknown users). To adapt them to our message-based
algorithms, we concatenate all tweets from each user, treating them as a
single message, as in~\cite{hong_discovering_2012}. Finally, the Eisenstein
data contain only unigrams from the text field (as we will show, including
information from other fields can notably improve results); for comparison, we
do the same. This yields 7,580 training and 1,895 test messages (i.e., roughly
380,000 tweets versus 13 million in our data set).

Judged by mean SAE, \algname{GMM-Opt-ID} surpasses all other approaches except
for Eisenstein et al.~\cite{eisenstein_sparse_2011}. Interestingly, the
algorithm ranking varies depending on whether mean or median SAE is used ---
e.g., \algname{GMM-Err-SAE16} has lower median SAE than
\cite{eisenstein_sparse_2011} but a higher mean SAE. This trade-off between
mean and median SAE also appears in other work -- for example,
Eisenstein~\etal{} report the best mean SAE but have much higher median
SAE~\cite{eisenstein_sparse_2011}. Also, Hong~\etal{} report the best median
SAE but do not report mean at all~\cite{hong_discovering_2012}.

Examining the results for \algname{GMM-Err-SAE} sheds light on this
discrepancy. We see that as the exponent increases from 4 to 16, the median
SAE decreases from 684\,km to 493\,km. However, calibration suffers rather
dramatically: \algname{GMM-Err-SAE16} has a quite overconfident $\oc{50} =
0.36$. This is explained in part by its use of fewer n-grams per message (182
for an exponent of 4 versus 37 for exponent 16).

Moreover, to our knowledge, no prior work reports either precision or
calibration metrics, making a complete comparison impossible. For example, the
better mean SAE of Eisenstein \etal~\cite{eisenstein_sparse_2011} may coincide
with worse precision or calibration. These metrics are not unique to our GMM
method, and we argue that they are critical to understanding techniques in
this space, as the trade-off above demonstrates.

Finally, we speculate that a modest decrease in accuracy may not outweigh the
simplicity and scalability of our approach. Specifically in contrast to topic
modeling approaches, our learning phase can be trivially parallelized by
n-gram.

\subsection{RQ2: Training size}

\xcaption{Accuracy of \algname{GMM-Err-SAE4} with different amounts of
  training data, along with the mean time to train and test one model.  Each
  day contains roughly 32,000 training tweets. (The 16-day test was run in a
  nonstandard configuration and its timing is therefore omitted.)}
\begin{figure}
  \includegraphics[width=\columnwidth]{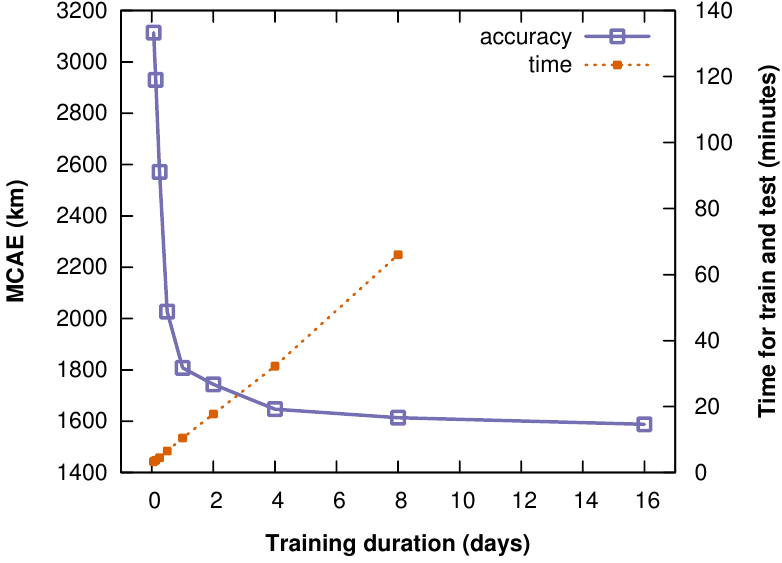}
  \vspace{\spacehackfig}
  \xcaptionplace
  \label{fig.training-size}
\end{figure}

We evaluated the accuracy of \algname{GMM-Err-SAE4} on different training
durations, no gap, all fields except user description, and minimum instances
of 3. We used a stride of 13 days for performance reasons.

Figure~\ref{fig.training-size} shows our results. The knee of the curve is
1~day of training (i.e., about 30,000 tweets), with error rapidly plateauing
and training time increasing as more data are added; accordingly, we use
1~training day in our other experiments.\footnote{We also observed
  deteriorating calibration beyond 1~day; this may explain some of the
  accuracy improvement and should be explored.}

\xcaption{Accuracy and run time of \algname{GMM-Err-SAE4} vs.\ inclusion
  thresholds for the number of times an n-gram appears in training data.}
\begin{figure}
  \includegraphics[width=\columnwidth]{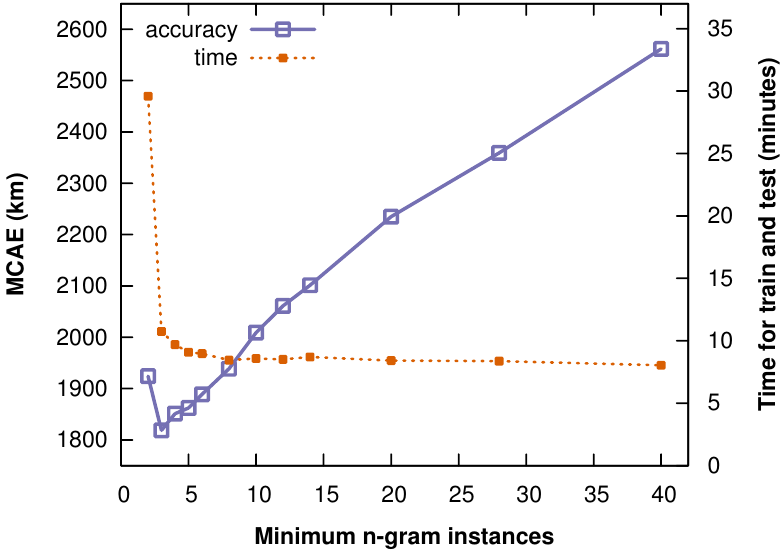}
  \vspace{\spacehackfig}
  \xcaptionplace
  \label{fig.min-instances}
\end{figure}

We also evaluated accuracy when varying minimum instances (the frequency
threshold for retaining n-grams), with training days fixed at 1;
Figure~\ref{fig.min-instances} shows the results. Notably, including n-grams
which appear only 3 times in the training set improves accuracy at modest time
cost (and thus we use this value in our other experiments). This might be
explained in part by the well-known long-tail distribution of word
frequencies; that is, while the informativeness of each individual n-gram may
be low, the fact that low-frequency words occur in so many tweets can impact
overall accuracy. This finding supports Wing \& Baldridge's
suggestion~\cite{wing_simple_2011} that
Eisenstein~\etal~\cite{eisenstein_latent_2010} pruned too aggressively by
setting this threshold to 40.

%

\subsection{RQ3: Time dependence}

\xcaption{Accuracy of GMM-Err-SAE4 with increasing delay between training and
  testing.}
\begin{figure}
  \includegraphics[width=\columnwidth]{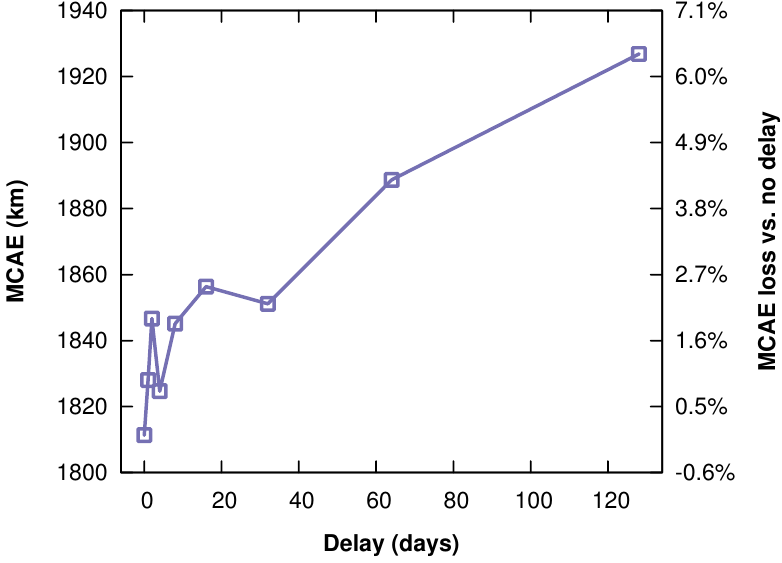}
  \vspace{\spacehackfig}
  \xcaptionplace
  \label{fig.gap}
\end{figure}

We evaluated the accuracy of \algname{GMM-Err-SAE4} on different temporal gaps
between training and testing, holding fixed training duration of 1~day and
minimum n-gram instances of 3. Figure~\ref{fig.gap} summarizes our results.
Location inference is surprisingly time-invariant: while error rises linearly
with gap duration, it does so slowly -- there is only about 6\% additional
error with a four-month gap. We speculate that this is simply because
location-informative n-grams which are time-dependent (e.g., those related to
a traveling music festival) are relatively rare.


\subsection{RQ4: Location signal source}
\label{sec.rq4}

We wanted to understand which types of content provide useful location
information under our algorithm. For example, Figure~\ref{fig.tweetwin} on the
first page illustrates a successful estimate by \algname{GMM-Err-SAE4}. Recall
that this was based almost entirely on the n-grams \token{angeles ca} and
\token{ca}, both from the location field. Table~\ref{tab.examples} in the
appendices provides a further snapshot of the algorithm's output. These hint
that, consistent with other methods (e.g.,~\cite{hecht_tweets_2011}), toponyms
provide the most important signals; below, we explore this hypothesis in more
detail.

\subsubsection{Which fields provide the most value?}
\label{sec.rq4.value}

One framing of this research question is structural. To measure this, we
evaluated \algname{GMM-Err-SAE4} on each combination of the five tweet fields,
holding fixed training duration at 1~day, gap at zero, and minimum instances
at 3. This requires an additional metric: \vocab{success rate} is the fraction
of test tweets for which the model can estimate a location (i.e., at least one
n-gram in the test tweet is present in the trained model).

%
%
\xcaption{Value of each field. \emph{Alone} shows the accuracy and success
  rate of estimation using that field alone, while \emph{Improvement} shows
  the mean improvement when adding a field to each combination of other fields
  (in both cases, positive indicates improvement). For example, adding user
  location to some combination of the other four fields will, on average,
  decrease MCAE by 1,255\,km and increase the success rate by 1.7 percentage
  points.}
\begin{table}
  \centering
  \tablefont
  
\begin{tabular}{@{}lrrrr@{}}
  \toprule
    \mc{1}{@{}c}{\mr{2}{*}{\textbf{Field}}}
  & \mc{2}{c}{\textbf{Alone}}
  & \mc{2}{c@{}}{\textbf{Improvement}}
  \\
  & \mc{1}{c}{\textbf{MCAE}}
  & \mc{1}{c}{\textbf{success}}
  & \mc{1}{c}{\textbf{MCAE}}
  & \mc{1}{c@{}}{\textbf{success}} \\
  \cmidrule(r){1-1}
  \cmidrule(lr){2-3}
  \cmidrule(l){4-5}
user location & 2125 & 65.8\% & 1255 & 1.7\% \\
user time zone & 2945 & 76.1\% & 910 & 3.0\% \\
tweet text & 3855 & 95.7\% & 610 & 7.3\% \\
user description & 4482 & 79.7\% & 221 & 3.3\% \\
user language & 6143 & 100.0\% & -103 & 8.5\% \\

  \bottomrule
\end{tabular}

  \vspace{\spacehacktab}
  \xcaptionplace
  \label{tab.fieldvalue}
\end{table}
\xcaption{Accuracy of including different fields. We list each combination of
  fields, ordered by increasing MCAE.}
\begin{table}
  \centering
  \tablefont
  
\begin{tabular}{@{}cccccccr@{}}
  \toprule
    \mc{1}{@{}c}{\textbf{Rank}}
  & \mc{5}{c}{\textbf{Fields}}
  & \mc{1}{c}{\textbf{MCAE}}
  & \mc{1}{c@{}}{\textbf{success}} \\
  \cmidrule(r){1-1}
  \cmidrule(lr){2-6}
  \cmidrule(l){7-8}
 1 & lo & tz & tx &    & ln & 1823 & 100.0\% \\
 2 & lo & tz & tx & ds & ln & 1826 & 100.0\% \\
 3 & lo & tz &    &    &    & 1862 & 87.7\% \\
 4 & lo & tz & tx &    &    & 1878 & 99.2\% \\
 5 & lo & tz & tx & ds &    & 1908 & 99.6\% \\
 6 & lo & tz &    & ds &    & 2013 & 94.1\% \\
 7 & lo & tz &    & ds & ln & 2121 & 100.0\% \\
 8 & lo &    &    &    &    & 2125 & 65.8\% \\
 9 & lo &    & tx & ds & ln & 2176 & 100.0\% \\
10 & lo & tz &    &    & ln & 2207 & 100.0\% \\
11 & lo &    & tx & ds &    & 2274 & 99.2\% \\
12 & lo &    & tx &    & ln & 2310 & 100.0\% \\
13 & lo &    & tx &    &    & 2383 & 98.0\% \\
14 &    & tz & tx & ds & ln & 2492 & 100.0\% \\
15 & lo &    &    & ds &    & 2585 & 88.3\% \\
16 &    & tz & tx & ds &    & 2594 & 99.4\% \\
17 &    & tz & tx &    & ln & 2617 & 100.0\% \\
18 &    & tz & tx &    &    & 2691 & 98.7\% \\
19 & lo &    &    & ds & ln & 2759 & 100.0\% \\
20 &    & tz &    &    &    & 2945 & 76.1\% \\
21 &    & tz &    & ds &    & 2991 & 91.8\% \\
22 &    & tz &    & ds & ln & 3039 & 100.0\% \\
23 & lo &    &    &    & ln & 3253 & 100.0\% \\
24 &    &    & tx & ds & ln & 3267 & 100.0\% \\
25 &    &    & tx & ds &    & 3426 & 98.8\% \\
26 &    & tz &    &    & ln & 3496 & 100.0\% \\
27 &    &    & tx &    & ln & 3685 & 100.0\% \\
28 &    &    & tx &    &    & 3855 & 95.7\% \\
29 &    &    &    & ds &    & 4482 & 79.7\% \\
30 &    &    &    & ds & ln & 4484 & 100.0\% \\
31 &    &    &    &    & ln & 6143 & 100.0\% \\

  \bottomrule

\end{tabular}

  \vspace{\spacehacktab}
  \xcaptionplace
  \label{tab.fieldtruth}
\end{table}

Table~\ref{tab.fieldvalue} summarizes our results, while
Table~\ref{tab.fieldtruth} enumerates each combination. User location and time
zone are the most accurate fields, with tweet text and language important for
success rate. For example, comparing the first and third rows of
Table~\ref{tab.fieldtruth}, we see that adding text and language fields to a
model that considers only location and timezone fields improves MCAE only
slightly (39\,km) but improves success rate considerably (by 12.3\% to
100.0\%). We speculate that while tweet text is a noisier source of evidence
than time zone (due to the greater diversity of locations associated with each
n-gram), our algorithm is able to combine these sources to increase both
accuracy and success rate.

It is also interesting to compare the variant considering only the location
field (row 8 of Table 4) with previous work that heuristically matches strings
from the location field to gazetteers. Hecht~\etal{} found that 66\% of user
profiles contain some type of geographic information in their location
field~\cite{hecht_tweets_2011}, which is comparable to the 67\% success rate
of our model using only location field.

Surprisingly, user description adds no value at all; we speculate
that it tends to be redundant with user location.

\subsubsection{Which types of n-grams provide the most value?}

We also approached this question by content analysis. To do so, from an
arbitrarily chosen test of the 58 successful \algname{GMM-Err-SAE4} tests, we
selected a ``good'' set of the 400 (or 20\%) lowest-CAE tweets, and a ``bad''
set of the 400 highest-CAE tweets. We further randomly subdivided these sets
into 100 training tweets (yielding 162 good n-grams and 457 bad ones) and 300
testing tweets (364 good n-grams and 1,306 bad ones, of which we randomly
selected 364).

Two raters independently created categories by examining n-grams from the
location and tweet text fields in the training sets. These were merged by
discussion into a unified hierarchy. The same raters then independently
categorized n-grams from the two fields into this hierarchy, using Wikipedia
to confirm potential toponyms and Google Translate for non-English n-grams.
Disagreements were again resolved by discussion.\footnote{We did a similar
  analysis of the language and time zone fields, using their well-defined
  vocabularies instead of human judgement. However, this did not yield
  significant results, so we omit it for brevity.}

%
\xcaption{Content analysis of n-grams in the location and text fields. For
  each category, we show the fraction of total weight in all location
  estimates from n-grams of that category; e.g., 49\% of all estimate weight
  in the good estimates was from n-grams with category \textit{city} (weights
  do not add up to 100\% because time zone and language fields are not
  included). Weights that are significantly greater in good estimates than bad
  (or vice versa) are indicated with a significance code (\px\ = 0.1, \pv\ =
  0.05, \poi\ = 0.01, \pooi\ = 0.001) determined using a Mann-Whitney U test
  with Bonferroni correction, the null hypothesis being that the mean weight
  assigned to a category over all n-grams in the \textit{good} set is equal to
  the mean weight for the same category in the \textit{bad} set. Categories
  with less than 1.5\% weight in both classes are rolled up into
  \textit{other}. We also show the top-weighted examples in each category.}
\begin{table*}
  \centering
  \begin{tabulary}{\textwidth}{@{}llrll@{}}
    \toprule
      \mc{2}{@{}c}{\textbf{Category}}
    & \mc{1}{c}{\textbf{Good}}
    & \mc{1}{c}{\textbf{Bad}}
    & \mc{1}{c@{}}{\textbf{Examples}} \\
    \cmidrule(r){1-2}
    \cmidrule(l){3-4}
    \cmidrule(l){5-5}
    \\[-6pt]
\textbf{location}	&	&	\textbf{\pooi\ 0.83}	&	\textbf{0.19\ }	&	\\
&	city	&	\pooi\ 0.49	&	0.09\ 	&	edinburgh, roma, leicester, houston tx	\\
&	country	&	\poi\ 0.10	&	0.03\ 	&	singapore, the netherlands, nederland, janeiro brasil	\\
&	generic	&	\ 0.01	&	0.02\ 	&	de mar, puerta de, beach, rd singapore	\\
&	state	&	\pooi\ 0.14	&	0.02\ 	&	maryland, houston tx, puebla, connecticut	\\
&	other lo	&	\pooi\ 0.09	&	0.02\ 	&	essex, south yorkshire, yorkshire, gloucestershire	\\
\textbf{not-location}	&	&	\textbf{\ 0.07}	&	\textbf{0.57\ \pooi}	&	\\
&	dutch word	&	\pooi\ 0.02	&	0.00\ 	&	zien, bij de, uur, vrij	\\
&	english word	&	\ 0.01	&	0.37\ \pooi	&	st new, i, pages, check my	\\
&	letter	&	\ 0.01	&	0.04\ 	&	\textmu, w, \textalpha, s	\\
&	slang	&	\ 0.00	&	0.08\ \pooi	&	bitch, lad, ass, cuz	\\
&	spanish word	&	\ 0.00	&	0.07\ \pooi	&	mucha, niña, los, suerte	\\
&	swedish word	&	\ 0.00	&	0.02\ 	&	rätt, jävla, på, kul	\\
&	turkish word	&	\ 0.02	&	0.00\ 	&	kar, restoran, biraz, daha	\\
&	untranslated	&	\ 0.02	&	0.00\ 	&	cewe, gading, ung, suria	\\
\textbf{technical}	&	&	\textbf{\poi\ 0.03}	&	\textbf{0.02\ }	&	\\
&	foursquare	&	\pooi\ 0.03	&	0.00\ 	&	paulo http, istanbul http, miami http, brasília http	\\
&	url	&	\ 0.00	&	0.02\ 	&	co, http, http t, co h	\\
\textbf{other}	&	&	\textbf{\ 0.03}	&	\textbf{0.04\ }	&	\\

    \\[-6pt]  
    \bottomrule
 \end{tabulary}
  \vspace{\spacehacktab}
  \xcaptionplace
  \label{tab.content}
\end{table*}

Our results are presented in Table~\ref{tab.content}. Indeed, toponyms offer
the strongest signal; fully 83\% of the n-gram weight in well-located tweets
is due to toponyms, including 49\% from city names. In contrast, n-grams used
for poorly-located tweets tended to be non-toponyms (57\%). Notably, languages
with geographically compact user bases, such as Dutch, also provided strong
signals even for non-toponyms.

These results and those in the previous section offer a key insight into
gazetteer-based
approaches~\cite{gelernter_geo-parsing_2011,hecht_tweets_2011,paradesi_geotagging_2011,schulz_multi-indicator_2013,valkanas_location_2012},
which favor accuracy over success rate by considering only toponyms. However,
our experiments show that both accuracy and success rate are improved by
adding non-toponyms, the latter to nearly 100\%; for example, compare rows~1
and~8 of Table~\ref{tab.fieldtruth}. Further, Table~\ref{tab.content} shows
that 17\% of location signal in well-located tweets is not from toponyms.

\section{Implications}
\label{sec.discussion}

We propose new judgement criteria for location estimates and specific metrics
to compute them. We also propose a simple, scalable method for location
inference that is competitive with more complex ones, and we validate this
approach using our new criteria on a dataset of tweets that is comprehensive
temporally, geographically, and linguistically.

This has implications for both location inference research as well as
applications which depend on such inference. In particular, our metrics can
help these and related inference domains better balance the trade-off between
precision and recall and to reason properly in the presence of uncertainty.

Our results also have implications for privacy. In particular, they suggest
that social Internet users wishing to maximize their location privacy should
(a)~mention toponyms only at state- or country-scale, or perhaps not at all,
(b)~not use languages with a small geographic footprint, and, for maximal
privacy, (c)~mention decoy locations. However, if widely adopted, these
measures will reduce the utility of Twitter and other social systems for
public-good uses such as disease surveillance and response. Our recommendation
is that system designers should provide guidance enabling their users to
thoughtfully balance these issues.

Future directions include exploring non-gaussian and non-parametric density
estimators and improved weighting algorithms (e.g., perhaps those optimizing
multiple metrics), as well as ways to combine our approach with others, in
order to take advantage of a broader set of location clues. We also plan to
incorporate priors such as population density and to compare with human
location assessments.

\section{Acknowledgments}

\blackout{Susan M. Mniszewski, Geoffrey Fairchild, and other members of our
  research team provided advice and support. We thank our anonymous reviewers
  for key guidance and the Twitter users whose content we studied. This work
  is supported by NIH/NIGMS/MIDAS, grant U01-GM097658-01. Computation was
  completed using Darwin, a cluster operated by CCS-7 at LANL and funded by
  the Accelerated Strategic Computing Program; we thank Ryan Braithwaite for
  his technical assistance.}
Maps were drawn using Quantum GIS;\footnote{\url{http://qgis.org}} base map
geodata is from Natural Earth.\footnote{\url{http://naturalearthdata.com}}
\blackout{LANL is operated by Los Alamos National Security, LLC for the
  Department of Energy under contract DE-AC52-06NA25396.}

\section{Appendix: Mathematical implementations}
\label{apx.math}

%

%
\xcaption{Example output of \algname{GMM-Err-SAE4} for an arbitrarily selected
  test. \vocab{TZ} is the time zone field (with \textit{-timeuscanada}
  omitted), while \vocab{L} is the language code. N-grams which collectively
  form 95\% of the estimate weight are listed. \vocab{CAE} is in kilometers,
  while \pra{\bm{50}} is in square kilometers.}
\begin{table*}
  \centering
  \small
  \begin{tabulary}{\textwidth}{@{}rLllclrr@{}}
    \toprule
      \mc{1}{@{}c}{\textbf{\%ile}}
    & \mc{1}{c}{\textbf{Tweet text}}
    & \mc{1}{c}{\textbf{Location}}
    & \mc{1}{c}{\textbf{TZ}}
    & \mc{1}{c}{\textbf{L}}
    & \mc{1}{c}{\textbf{N-grams}}
    & \mc{1}{c}{\textbf{CAE}}
    & \mc{1}{c@{}}{\textbf{\pra{\bm{50}}}} \\
    \midrule
    100 & I'm at Court Avenue Restaurant and Brewing Company (CABCO) (309 Court Avenue, Des Moines) w/ 3 others http://t.co/LW8cKUG3 & Urbandale, IA & central & en & \pbox[t]{10cm}{0.50 tx moines\vphantom{Hp} \\ 0.50 tx des moines\vphantom{Hp}} & 4 & 34 \\ 
\addlinespace
90 & Eyebrow threading time with \textit{@mention} :) & Cardiff , Wales &  & en & \pbox[t]{10cm}{0.73 lo cardiff\vphantom{Hp} \\ 0.27 lo wales\vphantom{Hp}} & 17 & 379 \\ 
\addlinespace
80 & Americans are optimistic about the economy \& like what Obama is doing. What is he doing? Campaigning and playing golf? Ignorance is bliss & Los Angeles, CA & pacific & en & \pbox[t]{10cm}{0.87 lo angeles ca\vphantom{Hp} \\ 0.12 lo ca\vphantom{Hp}} & 115 & 835 \\ 
\addlinespace
70 & Extreme Close Up.. & Rancagua, Chile & quito & es & \pbox[t]{10cm}{1.00 lo chile\vphantom{Hp}} & 272 & 1,517 \\ 
\addlinespace
60 & Reaksinya bakal sama ga yaa? Pengen tau.. http://t.co/8ABEPmKQ & ÜT: -2.9873722,104.7218631 & pacific & en & \pbox[t]{10cm}{0.97 tx pengen\vphantom{Hp}} & 451 & 2,974 \\ 
\addlinespace
50 & Follow \textit{@mention} exhibition date announced soon \#Fabulous & London  &  & en & \pbox[t]{10cm}{1.00 lo london\vphantom{Hp}} & 688 & 967 \\ 
\addlinespace
40 & You cannot you on ANY news station and NOT see NEWT being ripped apart. &  & quito & en & \pbox[t]{10cm}{0.99 tx newt\vphantom{Hp}} & 1,008 & 634,421 \\ 
\addlinespace
30 & \textit{@mention} kkkkkk besta &  & santiago & en & \pbox[t]{10cm}{0.91 tx kkkkkk\vphantom{Hp} \\ 0.08 tz santiago\vphantom{Hp}} & 1,496 & 511,405 \\ 
\addlinespace
20 & \textit{@mention} eu entrei no site é em dólar, se for real eu compro uma pra vc ir de novo Pra Disney agora. & Belem-PA & brasilia & pt & \pbox[t]{10cm}{0.89 tx de novo\vphantom{Hp} \\ 0.07 lo pa\vphantom{Hp}} & 2,645 & 263,576 \\ 
\addlinespace
10 & Þegar ég get ekki sofið \#hunangsmjolk http://t.co/zx43NoZD &  &  & en & \pbox[t]{10cm}{0.81 tx get\vphantom{Hp} \\ 0.05 ln en\vphantom{Hp} \\ 0.02 tx t\vphantom{Hp} \\ 0.02 tx zx\vphantom{Hp} \\ 0.02 tx co\vphantom{Hp} \\ 0.02 tx t co\vphantom{Hp}} & 5,505 & 2,185,354 \\ 
\addlinespace
0 & \textit{@mention} cyber creeping ya mean! I'm in New Zealand not OZ you mad \textit{expletive} haha it's deadly anyways won't b home anytime soon :P &  &  & en & \pbox[t]{10cm}{1.00 tx \textit{expletive}\vphantom{Hp}} & 18,578 & 17,827 \\ 

    \bottomrule
  \end{tabulary}
  \vspace{\spacehacktab}
  \xcaptionplace
  \label{tab.examples}
\end{table*}

\subsection{Metrics}
\label{apx.metrics}

This section details the mathematical implementation of the metrics presented
above. To do so, we use the following vocabulary. Let $m$ be a message
represented by a binary feature vector of n-grams (i.e., sequences of up to
$n$ adjacent tokens; we use $n=2$) $m=\{w_1\ldots w_V\}$, $w_j \in \{0,1\}$.
$w_j=1$ means that n-gram $w_j$ appears in message $m$, and $V$ is the total
size of the vocabulary. Let $y \in \mathbbm{R}^2$ represent a geographic point
(for example, latitude and longitude) somewhere on the surface of the Earth.
We represent the true origin of a message as $y^*$; given a new message $m$,
our goal is to construct a geographic density estimate $f(y|m)$, a function
which estimates the probability of each point $y$ being the true origin of
$m$.

These implementations are valid for any density estimate $f$, not just
gaussian mixture models. Specific types of estimates may require further
detail; for GMMs, this is noted below.

CAE depends further on the geodesic distance $d(y,y^*)$ between the true
origin $y^*$ and some other point $y$. It can be expressed as:
\begin{align}
  \label{eq:cae}
  \text{CAE} = \text{E}_f [d(y,y^*)] = \int_y d(y,y^*)f(y|m) \dy
\end{align}

As computing this integral is intractable in general, we approximate it using
a simple Monte Carlo procedure. First, we generate a random sample of $n$
points from the density $f$, $S=\{y_1 \ldots y_n\}$ ($n=1000$ in our
experiments).\footnote{The implementations of our metrics depend on being able
  to efficiently (a)~sample a point from $f$ and (b)~evaluate the probability
  of any point.} Using this sample, we compute CAE as follows:
\begin{equation}
  \label{eq:cae.approx}
  \text{CAE} \approx \frac{1}{|S|} \sum_{y \in S} d(y,y^*)
\end{equation}

Note that in this implementation, the weighting has become implicit: points
that are more likely according to $f$ are simply more likely to appear in
$S$. Thus, if $f$ is a good estimate, most of the samples in $S$ will be near
the true origin.

To implement PRA, let $R_{f,\beta}$ be a prediction region such that the
probability of $y^*$ falling within the geographic region $R$ is its coverage
$\beta$. Then, \pra{\beta} is simply the area of $R$:

\begin{equation}
  \label{eq:pra}
  \pra{\beta} = \int_{R_{f,\beta}}\!\!\dy
\end{equation}

As above, we can use a sample of points $S$ from $f$ to construct an
approximate version of $R$:

\begin{enumerate}

\item Sort $S$ in descending order of likelihood $f(y_i|m)$. Let $S_\beta$ be
  the set containing the top $|S|\beta$ sample points.

\item Divide $S_\beta$ into approximately convex clusters.

\item For each cluster of points, compute its convex hull, producing a
  geo-polygon.

\item The union of these hulls is approximately $R_{f,\beta}$, and the area of
  this set of polygons is approximately \pra{\beta}.\footnote{Because the
    polygons lie on an ellipsoidal Earth, not a plane, we must compute the
    \vocab{geodesic area} rather than a planar area. This is accomplished by
    projecting the polygons to the Mollweide equal-area projection and
    computing the planar area under that projection.}

\end{enumerate}

Finally, recall that \oc{\beta} for a given estimator and a set of test
messages is the fraction of tests where $y^*$ was within the prediction region
$R_{f,\beta}$. That is, for a set $(y_1^*, y_2^*, ... y_n^*)$ of $n$ true
message origins:

\begin{equation}
  \label{eq:oc}
  \oc{\beta} = \frac{1}{n} \sum_{i=1}^n \mathbbm{1}[y_i^* \in R^i_{f,\beta}]
\end{equation}

We do not explicitly test whether $y^* \in R$, because doing so propagates any
errors in approximating $R$. Instead, we count how many samples in $S$ have
likelihood less than $f(y^*|m)$; if this fraction is greater than $\beta$,
then $y^*$ is (probably) in $R$. Specifically:
\begin{equation}
  \label{eq:fraction}
  r(y^*) = \frac{1}{|S|} \sum_{y \in S} \mathbbm{1}[f(y) < f(y^*)]
\end{equation}
\begin{equation}
  \label{eq:oc2}
  \oc{\beta} \approx \frac{1}{n} \sum_{i=1}^n \mathbbm{1}[r(y_i^*) > \beta]
\end{equation}

\subsection{Gaussian mixture models}
\label{apx.gmms}

As introduced in section ``Our Approach'', we construct our location model by
training on geographic data consisting of a set $D$ of $n$ (message, true
origin) pairs extracted from our database of geotagged tweets; i.e., $D =
\{(m_i, y^*_i)\}_{i=1}^{n}$. For each n-gram $w_j$, we fit a gaussian mixture
model $g(y|w_j)$ based on examples in $D$. Then, to estimate the origin
location of a new message $m$, we combine the mixture models for all n-grams
in $m$ into a new density $f(y|m)$. These steps are detailed below.

We estimate $g$ for each (sufficiently frequent) n-gram $w_j$ in $D$ as
follows. First, we gather the set of true origins of all messages containing
$w_j$, and then we fit a gaussian mixture model of $r$ components to represent
the density of these points:
\begin{equation}
  g(y|w_j) = \sum_{k=1}^r \pi_k^j \, \mathcal{N}(y|\mu_k^j, S_k^j)
\end{equation}
where $\pi^j = \{\pi^j_1 \ldots \pi^j_r\}$ is a vector of mixture weights and
$\mathcal{N}$ is the normal density function with mean $\mu^j_k$ and
covariance $S^j_k$. We refer to $g(y|w_j)$ as an \vocab{n-gram density}.

We fit $\pi$ and $S$ independently for each n-gram using the expectation
maximization algorithm, as implemented in the Python package
scikit-learn~\cite{pedregosa_scikit-learn:_2011}.

Choosing the number of components $r$ is a well-studied problem. While
Dirichlet process mixtures~\cite{neal_markov_2000} are a common solution, they
can scale poorly. For simplicity, we instead investigated a number of
heuristic approaches from the literature~\cite{milligan_examination_1985}; in
our case, $r = \min(m, \log(n)/2)$ worked well, where $n$ is the number of
points to be clustered, and $m$ is a parameter. We use this heuristic with
$m=20$ in all experiments.

Next, to estimate the origin of a new message $m$, we gather the available
densities $g$ for each n-gram in $m$ (i.e., some n-grams may appear in $m$ but
not in sufficient quantity in $D$). We combine these n-gram densities into a
mixture of GMMs:
\begin{align}
  \label{eq:mdensity}
  f(y|m) &= \sum_{w_j \in m} \delta_j g(y|w_j) = \sum_{w_j \in m} \delta_j
  \sum_{k=1}^r \pi_k^j \, \mathcal{N}(y|\mu_k^j, S_k^j)
\end{align}
where $\delta=\{\delta_1 \ldots \delta_V\}$ are the \vocab{n-gram mixture
  weights} associated with each n-gram density $g$. We refer to $f(y|m)$ as a
\vocab{message density}.

A mixture of GMMs can be implemented as a single GMM by multiplying $\delta_j$
by $\pi^j_k$ for all $j,k$ and renormalizing so that the mixture weights sum
to 1. Thus, Equation \ref{eq:mdensity} can be rewritten:
\begin{align}
\label{eq:mdensity.norm}
f(y|m) = \sum_{w_j \in m}
         \sum_{k=1}^r \tau_k^j
         \, \mathcal{N}(y|\mu_k^j, S_k^j)
\end{align}
where $\tau_k^j = \delta_j \pi_k^j / \sum_{j,k} \delta_j \pi_k^j$.

We can now compute all four metrics. CAE and \oc{\beta} require no additional
treatment. To compute SAE, we distill $f(y|m)$ into a single point estimate by
the weighted average of its component means: $\hat{y}=\sum_{w_j \in m}
\sum_{k=1}^r \tau_k^j \mu_k^j$. Computing \pra{\beta} requires dividing
$S_\beta$ into convex clusters; we do so by assigning each point in $S$ to its
most probable gaussian in $f$.

The next two sections describe methods to set the n-gram mixture weights
$\delta_j$.

\subsection{Setting $\delta_j$ weights by inverse error}

Mathematically, the inverse error approach introduced above can be framed as a
non-iterative optimization problem. Specifically, we set $\delta$ by fitting a
multinomial distribution to the observed error distribution. Let $e_{ij} \in
\mathbbm{R}_{\ge 0}$ be the error incurred by n-gram density $g(y|w_j)$ for
message $m_i$; in our implementation, we use SAE as $e_{ij}$ for performance
reasons (results with CAE are comparable). Let $e_j$ be the average error of
n-gram $w_j$: $e_j = \frac{1}{N_j}\sum_{i=1}^{N_j} e_{ij}$, where $N_j$ is the
number of messages containing $w_j$. We introduce a model parameter $\alpha$,
which places a non-linear (exponential) penalty on error terms $e_j$. The
problem is to minimize the negative log likelihood, with constraints that
ensure $\delta$ is a probability distribution:
\begin{align}
\label{eq:inv.sae}
\delta^* \leftarrow & \argmin_\delta -\log \prod_j \delta_j^{\frac{1}{e_j^\alpha}}\\
\label{eq:inv.sae2}
& \text{s.t. } \sum_j \delta_j = 1 \text{ and } \delta_j \ge 0 \text{ } \forall j
\end{align}

This objective can be minimized analytically. While the inequality constraints
in Equation~\ref{eq:inv.sae2} will be satisfied implicitly, we express the
equality constraints using a Lagrangian:
\begin{align}
L(\delta, \lambda) &= -\log \prod_j \delta_j^{\frac{1}{e_j^\alpha}} + \lambda \left(\sum_j \delta_j - 1 \right)\\
&= -\sum_j \frac{1}{e_j^\alpha} \log \delta_j + \lambda \left(\sum_j \delta_j -1 \right)
\end{align}

Taking the partial derivative with respect to $\delta_k$ and setting to 0
results in:
\begin{align}
\label{eq.deriv}
\frac{\partial L}{\partial \delta_k} & = -\frac{1}{e_k^\alpha \delta_k} + \lambda = 0 \text{ }
\forall k\\
\label{eq.deriv2}
& = -\frac{1}{e_k^\alpha} + \lambda \delta_k = 0 \text{ } \forall k \\
\label{eq.deriv3}
\frac{\partial L}{\partial \delta} & = -\sum_k \frac{1}{e_k^\alpha} + \lambda \sum_k \delta_k = 0
\end{align}

The equality constraint lets us substitute $\sum_k \delta_k = 1$ in
Equation~\ref{eq.deriv3}. Solving for $\lambda$ yields:
\begin{equation}
  \lambda = \sum_k \frac{1}{e_k^\alpha}
\end{equation}

Plugging this into \ref{eq.deriv} and solving for $\delta_k$ results in:
\begin{equation}
  \delta_k = \frac{\frac{1}{e_k^\alpha}}{\sum_k\frac{1}{e_k^\alpha}}
\end{equation}

This brings us full circle to the intuitive result above: that the weight of an
n-gram is proportional to its average error.\footnote{Our implementation first
  assigns $\delta_k=\frac{1}{e_k^\alpha}$, then normalizes the weights
  per-message as in Equation~\ref{eq:mdensity.norm}.}

\subsection{Setting $\delta_j$ weights by optimization}

This section details the data-driven optimization algorithm introduced above.
We tag each n-gram density function with a feature vector. This vector
contains the ID of the n-gram density function, the quality properties, or
both of these. For example, the feature vector for the n-gram \textit{dallas}
might be $\{\text{id}=1234, \text{variance}=0.56, \text{BIC}=0.01, ...\}$. We
denote the feature vector for n-gram $w_j$ as $\phi(w_j)$, with elements
$\phi_k(w_j) \in \phi(w_j)$.

This feature vector is paired with a corresponding real-valued parameter
vector $\theta=\{\theta_1, \ldots, \theta_p\}$ setting the weight of each
feature. The vectors $\theta$ and $\phi$ are passed through the logistic
function to ensure the final weights $\delta$ are in the interval [0,1]:
\begin{align}
  \label{eq:delta}
  \delta_j^\theta = \frac{1}{1 + e^{-\sum_{k=1}^p \phi_k(w_j)\theta_k}}
\end{align}

The goal of this approach is to assign values to $\theta$ such that properties
that are predictive of low-error n-grams have high weight (equivalently, so
that these n-grams have large $\delta_j^\theta$). This is accomplished by
minimizing an error function (built atop the same SAE-based $e_{ij}$ as the
previous method):
\begin{align}
  \label{eq:opt.error}
  \theta^* & \leftarrow \argmin_\theta \sum_{i=1}^{|D|} \frac{\sum_{w_j \in m_i} e_{ij}
    \delta_j^\theta}{\sum_{w_j \in m_i} \delta_j^\theta}
\end{align}

After optimizing $\theta$, we assign $\delta^* = \delta^{\theta^*}\!$. The
numerator in Equation \ref{eq:opt.error} computes the sum of mixture weights
for each n-gram density weighted by its error; the denominator sums mixture
weights to ensure that the objective function is not trivially minimized by
setting $\delta_j^\theta$ to 0 for all $j$. Thus, to minimize Equation
\ref{eq:opt.error}, n-gram densities with large errors must be assigned small
mixture weights.

Before minimizing, we first augment the error function in Equation
\ref{eq:opt.error} with a regularization term:
\begin{align}
  \label{eq:error}
  \Phi(D,\theta) = \sum_{i=1}^{|D|} \frac{\sum_{w_j \in m_i} e_{ij}
    \delta_j^\theta}{\sum_{w_j \in m_i} \delta_j^\theta} + \frac{\lambda}{2} \|\theta\|^2
\end{align}

The extra term is an $\ell_2$-regularizer to encourage small values of
$\theta$ to reduce overfitting; we set $\lambda=1$ in our
experiments.\footnote{$\lambda$ could be tuned on validation data; this should
  be explored.}

We minimize Equation~\ref{eq:error} using gradient descent. For brevity, let
$n_{ij} = \sum_{w_j \in m_i} e_{ij} \delta_j^\theta$ and $d_{ij}=\sum_{w_j \in
  m_i}\delta_j^\theta$ be the numerator and denominator terms from Equation
\ref{eq:error}. Then, the gradient of Equation \ref{eq:error} with respect to
$\theta_k$ is
\begin{equation}
  \label{eq:gradient}
    \frac{\partial \Phi}{\partial \theta_k} =
    \sum_{i=1}^{|D|}\sum_{w_j \in
      m_i}\frac{-\phi_k(w_j)\delta_j^\theta(1-\delta_j^\theta) (e_{ij}d_{ij} - n_{ij})}{d_{ij}^2}
    + \lambda \theta_k
\end{equation}

We set Equation \ref{eq:gradient} to 0 and solve for $\theta$ using L-BFGS as
implemented in the SciPy Python package~\cite{jones_scipy:_2001}. (Note that
by decomposing the objective function by n-grams, we need only compute the
error metrics $e_{ij}$ once prior to optimization.) Once $\theta$ is set, we
then find $\delta$ according to Equation~\ref{eq:delta} and use these values
to find the message density in Equation~\ref{eq:mdensity}.

\section{Appendix: Tokenization algorithm}
\label{app.tokenization}

This section details our algorithm to convert a text string into a sequence of
n-grams, used to tokenize the message text, user description, and user
location fields into bigrams (i.e., $n=2$).

\begin{enumerate}

\item Split the string into candidate tokens, each consisting of a sequence of
  characters with the same Unicode category and script.
  Candidates not of the \vocab{letter} category are discarded, and letters are
  converted to lower-case. For example, the string ``Can't wait for
  \begin{CJK}{UTF8}{min}私の\end{CJK}'' becomes five
  candidate tokens: \emph{can}, \emph{t}, \emph{wait}, \emph{for}, and
  \begin{CJK}{UTF8}{min}私の\end{CJK}.

\item Candidates in certain scripts are discarded either because they do not
  separate words with a delimiter (Thai, Lao, Khmer, and Myanmar, all of which
  have very low usage on Twitter) or may not really be letters (Common,
  Inherited). Such scripts pose tokenization difficulties which we leave for
  future work.

\item Candidates in the scripts Han, Hiragana, and Katakana are assumed to be
  Japanese and are further subdivided using the TinySegmenter
  algorithm~\cite{hagiwara_tinysegmenter_????}. (We ignore the possibility
  that text in these scripts might be Chinese, because that language has very
  low usage on Twitter.) This step would split
  \begin{CJK}{UTF8}{min}私の\end{CJK} into
  \begin{CJK}{UTF8}{min}私\end{CJK} and
  \begin{CJK}{UTF8}{min}の\end{CJK}.

\item Create n-grams from adjacent tokens. Thus, the final tokenization of the
  example for $n=2$ would be: \emph{can}, \emph{t}, \emph{wait}, \emph{for},
  \begin{CJK}{UTF8}{min}私\end{CJK},
  \begin{CJK}{UTF8}{min}の\end{CJK}, \emph{can t}, \emph{t wait},
  \emph{wait for}, \emph{for} \begin{CJK}{UTF8}{min}私\end{CJK}, and
  \begin{CJK}{UTF8}{min}私\end{CJK}
  \begin{CJK}{UTF8}{min}の\end{CJK}.

\end{enumerate}

\section{Appendix: Results of pilot experiments}
\label{app.pilots}

This section describes briefly three directions we explored but did not pursue
in detail because they seemed to be of limited potential value.

\begin{itemize}

\item \inhead{Unifying fields.} Ignoring field boundaries slightly reduced
  accuracy, so we maintain these boundaries (i.e., the same n-gram appearing
  in different fields is treated as multiple, separate n-grams).

\item \inhead{Head trim.} We tried sorting n-grams by frequency and removing
  various fractions of the most frequent n-grams. In some cases, this yielded
  a slightly better MCAE but also slightly reduced the success rate;
  therefore, we retain common n-grams.

\item \inhead{Map projection.} We tried plate carr\'{e}e (i.e., WGS84
  longitude and latitude used as planar X and Y coordinates), Miller, and
  Mollweide projections. We found no consistent difference with our error- and
  optimization-based algorithms, though some others displayed variation in
  MPRA. Because this did not affect our results, we used plate carr\'{e}e for
  all experiments, but future work should explore exactly when and why map
  projection matters.

\end{itemize}


\shrunkenfont
\bibliographystyle{abbrv}
\bibliography{refs-cooked}

\end{document}